\documentstyle[12pt]{article}
\def\np    { Nucl. Phys. }

\def\pl    { Phys. Lett. }

\def\beqa{\begin{eqnarray}}
\def\eeqa{\end{eqnarray}}
\def\parn              {  \par\noindent }
\def\parbigskip        {  \par\bigskip  }

\def\parbigskipn        {  \par\bigskip\noindent  }

\def\papertitlepage{\baselineskip 3.5ex \thispagestyle{empty}}
\def\Title#1{\vspace{1.5cm}\begin{center}
 {\large
\bf #1} \end{center} 
\vspace{1cm}}
\def\Authors#1{\begin{center} {\large
\it #1} \end{center}}
\def\Abstract{\vspace{0.3cm}\begin{center} {\large
\bf Abstract} 
           \end{center} \parbigskip}
\def\ICRRnumber#1#2#3{\hfill \begin{minipage}{3cm} #1
              \parn #2 \parn #3 \end{minipage}}
\begin{document}
\papertitlepage
\vspace*{-1 cm}
\ICRRnumber{ }{March 2000}{hep-th/0003240}
\Title{On dimensional reductions of the M-9-brane }
\Authors{{\sc\  Takeshi Sato
\footnote{tsato@icrr.u-tokyo.ac.jp}} \\
 \vskip 3ex
 Institute for Cosmic Ray Research, University of Tokyo, \\
5-1-5 Kashiwa-no-ha, 
Kashiwa, Chiba, 277-8582  Japan \\
}
\Abstract

We discuss 
the relations of the M-9-brane with
other branes via dimensional reductions, 
mainly focusing on their Wess-Zumino (WZ) actions. 
It is shown that on three kinds of dimensional reductions, 
the WZ action of the M-9-brane respectively 
gives those of the D-8-brane, the ``KK-8A brane'' 
(which we regard as a kind of D-8-brane)
and the ``NS-9A brane'',
the last two actions of which were obtained via dualities. 
Based on these results,
we conclude that the relation of p-branes for $p\ge 8$,
proposed previously, is consistent
from the viewpoint of worldvolume actions.

\newpage

\section{Introduction}
M-theory is a candidate for a unified theory of particle 
interactions
and is conjectured to be the 11-dimensional (11D) theory
\cite{tow2}\cite{wit1}
which gives 5 perturbative 10-dimensional (10D) string theories 
in different kinds of limits. 
In discussing properties of these theories,
(p+1)-dimensional objects, called p-branes, 
play many crucial roles (e.g. \cite{pol2}), 
so, it is important to clarify
what kinds of branes exist in each of the theories.
Brane scan via superalgebra is one of the methods to discuss them,
in which one can read BPS branes possible to exist in the theories
from the structure of central charges of 
superalgebras\cite{hullalg}\cite{towalg}. 
For $p \le 7$, all of the BPS branes predicted from superalgebras to
exist in M-, IIA and IIB string theories 
have corresponding solutions in each of the supergravities 
i.e. the low energy limits of the theories. 
However, the p-branes we want to discuss in this paper are 
those with $p\ge 8$, for which 
one kind of 9-brane is suggested to exist in M-theory\cite{hullalg}, 
one kinds of 8-brane and 9-branes are predicted in
IIA, and two kinds of 9-branes are in IIB\cite{hullalg}.
The first one is called "M-9-brane", and the others are called 
(or identified with) 
D-8-brane, NS-9A-brane, D-9-brane and NS-9B-brane,
respectively, based on the consideration of the kinds of charges they
are supposed to have\cite{hullalg}.
Taking into account the dimensions and the duality relations of
the theories, the relation of the p-branes for 
$p\ge 8$, suggested from superalgebras, 
are represented as 
Figure 1\cite{hullalg} (see also \cite{hulm9}).
(``KK8A'' will be explained below.)
These branes are very important in that 
M- and string theories with 16 supercharges 
are expected to be constructed by using these branes\cite{pol2}%
\cite{pol1}\cite{pol3}\cite{hullalg}\cite{hulso32}\cite{sptfilling}
(see Figure 2).\footnote{``Horava-Witten'' in Figure 2 
denotes the Horava-Witten construction of M-theory\cite{horavawitten}.}
We will discuss these branes.

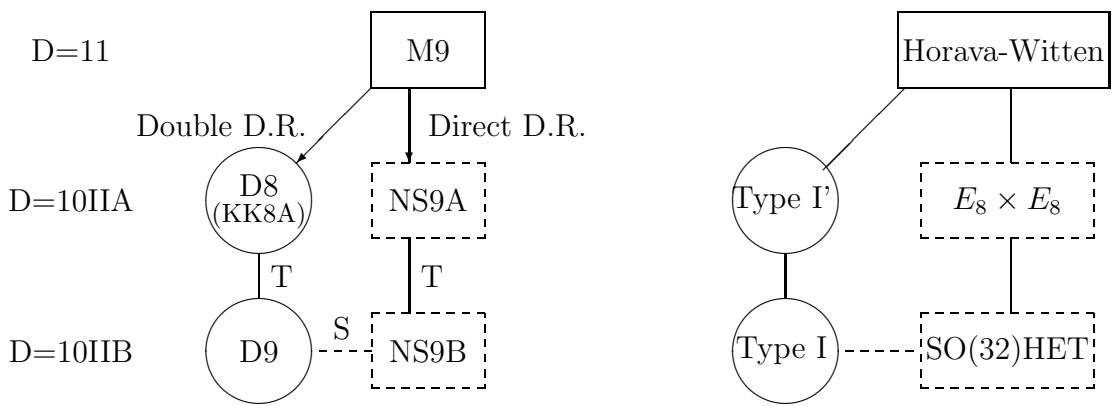
\begin{figure}[h]
 \begin{center}
  \setlength{\unitlength}{1mm}
\begin{picture}(110,55)
\put(-20,45){\makebox(20,10){D=11}}
\put(-20,25){\makebox(20,10){D=10IIA}}
\put(-20,5){\makebox(20,10){D=10IIB}}
\put(30,25){\dashbox(15,10){NS9A}}
\put(35,45){\vector(0,-10){10}}
\put(30,45){\framebox(15,10){M9}}
\put(30,45){\vector(-1,-1){10}}
\put(15,30){\circle{16}}
\put(10,28){\makebox(10,8){D8}}
\put(10,24){\makebox(10,8){{\footnotesize (KK8A)}}}
\put(35,25){\line(0,-10){10}}
\put(15,23){\line(0,-10){6}}
\put(30,5){\dashbox(15,10){NS9B}}
\put(15,10){\circle{16}}
\put(10,5){\makebox(10,10){D9}}
\put(32,15){\makebox(12,10){T}}
\put(12,15){\makebox(12,10){T}}
\multiput(23,10)(2,0){4}{\line(1,0){1}}
\put(20,8){\makebox(12,10){S}}
\put(43,35){\makebox(10,10){Direct D.R.}}
\put(5,35){\makebox(10,10){Double D.R.}}
\put(103,25){\dashbox(23,10){$E_{8}\times E_{8}$}}
\put(115,45){\line(0,-10){10}}
\put(100,45){\framebox(28,10){Horava-Witten}}
\put(101,45){\line(-1,-1){11}}
\put(85,30){\circle{16}}
\put(73,25){\makebox(23,10){Type I'}}
\put(115,25){\line(0,-10){10}}
\put(85,23){\line(0,-10){6}}
\put(103,5){\dashbox(23,10){\ SO(32)HET }}
\put(85,10){\circle{16}}
\put(73,5){\makebox(23,10){Type I}}
\multiput(93,10)(2,0){5}{\line(1,0){1}}
\end{picture}
 \end{center}
  \caption{Relations of p-branes with $p \ge 8$
\ \ \ Figure 2: M and strings with 16 SUSY}
  \label{fig:gen1}
\end{figure}

However, before presenting our work, we have to mention
the following problem with the M-9-brane, or 11D origins of
the D-8-brane solution and massive IIA SUGRA, and related issues.
The BPS D-8-brane arising in the IIA string, 
actually, has a corresponding solution not in the usual IIA 
but in the massive IIA SUGRA
with non-vanishing cosmological 
term\cite{pol2}\cite{pol1}\cite{berg3}. 
This is because a BPS D-8-brane in 10 dimensions is a domain wall 
with some electric charge of a RR 9-form gauge field, 
giving rise to a constant field strength, which we denote as a mass
parameter $m$ in this paper. 
This field strength contributes to
the action as the cosmological constant $-m^{2}/2$.\footnote{
In addition, the mass term of the NSNS 2-form B arises in the action
of IIA SUGRA in this case\cite{rom1}\cite{berg3}, 
so, the theory is called ``massive'' theory
and the corresponding background is called massive background.} 
In other words, such domain wall
solutions cannot be constructed without cosmological term. In 11D, 
however,
no deformation to include a cosmological term
is allowed if Riemannian
geometry and covariant action are assumed\cite{des1}. 
Thus, there is no naive M-9-brane
solution in 11D, and the origin of the D-8-brane and massive IIA SUGRA
are still unclear.

There are several approaches to solve 
this problem\cite{hlw1}\cite{llp1}\cite{berg4}\cite{hull3},
and one of them is "massive 11D theory"\cite{berg4}.
This is a trial theory, constructed
on the basis of the idea that {\it the
problem may imply the need to modify the framework of 11D SUGRA}.
Suppose a Killing isometry is assumed in the 11D background. Then, 
the no-go theorem is avoided and the massive 11D theory, which is
written in terms of an 11-dimensional theory at least formally,
can be defined;
it gives the 10D massive IIA SUGRA on dimensional reduction 
along the isometry direction 
(which is parametrized by the coordinate $z$), 
and gives usual 11D SUGRA in the massless limit $m \to 0$
if the dependence of the fields on $z$ is restored.
Moreover, the M-9-brane solution, i.e. the solution 
which gives a D-8-brane solution on the dimensional 
reduction along $z$, is obtained in this theory\cite{berg3}
(see also \cite{satom9}).
We note that only the bosonic sectors have been discussed in this
massive 11D theory, though its bulk theory is called ``super''gravity.
We also note that the isometry direction is interpreted
as a compactified direction like $S^{1}$, and 
the M-9-brane is considered to
be wrapped around it\cite{berg3} 
(see also \cite{hulm9}).\footnote{
More detailed target-space picture of the M-9-branes have been
discussed in ref.\cite{bergm9}\cite{sptfilling}. 
The tension of the M-9-brane is discussed in 
ref.\cite{hulm9}\cite{bergm9}.
For K-theoretic discussion on 
M-9-brane, see ref.\cite{ktheory}.
For brane decent relations including the M-9-brane, see 
ref.\cite{lozdecent}.}
We follow the above idea and 
study the relations of branes within this framework.

Then, there arises a further issue on dimensional reductions of the
M-9-brane, due to the existence of the isometry;
if one dimensionally reduce the M-9-brane along 
the worldvolume direction different from the isometry,
it is considered to give an 8-brane 
with one isometry direction as a worldvolume one. 
This brane 
is called 
``KK-8A brane'', whose properties have been discussed 
in ref.\cite{meessen1}\cite{eyras1}\cite{exotic} 
(see also \cite{hulm9}):
The KK-8A brane is described as a solution of another massive 
IIA SUGRA which is obtained by reducing the massive 11D SUGRA 
in a direction
different from the isometry one\cite{meessen1},
and the KK-8A brane is conjectured to be the T-dual of the ``NS-7B
brane'' which is the S-dual of the D-7-brane.\cite{meessen1}%
\cite{eyras1}\cite{exotic}.
The problem is, {\it to what extent we
should take it as an object independent of the D-8-brane};
if one considers the introduction of the Killing vector to be 
some intrinsic modification of the theory and
regards the isometry direction as a certain special one,
this is surely the third possibility of the dimensional reduction 
of the M-9-brane. 
On the other hand,
if the isometry originates from some M-theory 
background (for example, the one considered by Hull 
in ref.\cite{hull3}), it is not the third possibility
and the KK-8A brane solution
should be regarded as a kind of description of the D-8-brane
in some special background.
In ref.\cite{exotic} 
it is argued that 
from the viewpoint of target-space solutions,
the two 8-branes represent the same physical object
since the solutions relate to each other via a certain coordinate
transformation\footnote{See (the pages 33 and 34 in section 6 of the
hep-th version of) ref.\cite{exotic}.}
(although from the viewpoint of spacetime superalgebra, 
the KK-8A brane should be different from the D-8-brane).
Based on this argument, we consider that the isometry
emerges from some background, as discussed in ref.\cite{sptfilling},
and discuss the KK-8A brane as a kind of D-8-brane.
(In addition, we will give a certain
comment on the relation of the KK-8A brane and
superalgebras in the final section.)

To be concrete, we will discuss the relations from the viewpoint of
worldvolume effective actions (WVEAs). 
In fact, all the WVEAs of the bosonic
sectors of branes have been 
constructed\cite{massDbr1}\cite{massDbr2}%
\cite{massDbr3}\cite{berg4}%
\cite{loz1}\cite{bergm9}\cite{lozkkm}\cite{eyras1}%
\cite{exotic}\cite{satom9}.
In particular,
the worldvolume actions of the NS-9A brane and 
the KK-8A brane (as a D-8-brane)
have been obtained 
{\it via chains of dualities}.\footnote{
To be concrete, 
the action of 
the KK-8A brane has been obtained from that of 
the D-7-brane by S- and T-dualizing it\cite{exotic}.} 
However, 
the relations of the action of the M-9-brane with 
those of the D-8-brane and the NS-9A brane 
{\it via dimensional reductions}
have not been discussed for the full action;
the discussion were done 
only for their kinetic actions\cite{eyras1},
merely because the WZ action of the M-9-brane has been
obtained only recently\cite{satom9}.\footnote{The relation 
of the WZ action of the M-9-brane with
that of the D-8-brane was discussed partly in ref.\cite{satom9}, 
only for the case of the dimensional reduction along $z$.}
In other words, {\it even within this framework, 
the consistency of the
relation of the branes has not been established yet}.

The purpose of this paper is to
examine the consistency of the relations of the p-branes 
with $p \ge 8$ in Fig.1
from the viewpoint of WVEAs, 
by discussing the dimensional reductions of the  M-9-brane WZ action 
and comparing them with the WZ actions 
of the D-8-brane and the NS-9A brane.

The organization of this paper is as follows:
In section 2, we first review 
the massive 11D SUGRA and construction
of the M-9-brane WZ action.
A certain logical mistake in ref.\cite{satom9} 
in constructing it is corrected.
(To be concrete, the action should be constructed only on the basis 
of its gauge invariance, but should 
not be so constructed to give 
the D-8-brane WZ action on dimensional reduction.)
In section 3 
we discuss three kinds of dimensional reductions of the action,
and compare them with the actions of the D-8-brane, 
the KK-8A brane and the NS-9A brane, respectively.
In section 4 we give summary and discussion,
especially on the massive part of the KK-8A brane WZ action, 
and the relation of the KK-8A branes with spacetime superalgebras.

The notation of this paper is as follows:
We use ``mostly-minus'' metrics for both target-spaces and
worldvolumes. 
Fields, indices and coordinates
with hats are 11-dimensional, 
while those with no hats are 10-dimensional.
We denote spacetime coordinates by 
$\hat{x}^{\hat{\mu}},\hat{x}^{\hat{\nu}},\cdots$
or $x^{\mu},x^{\nu},\cdots$,
and local Lorentz indices by $\hat{a},\hat{b},\cdots$ or $a,b,\cdots$.
Finally, we set $2\pi \alpha'=1$.


\section{The massive 11D SUGRA and the M-9-brane WZ action}
\setcounter{footnote}{0} 


\subsection{Review of the massive 11D SUGRA}

In this section we first review the massive 11D
supergravity\cite{berg4}\cite{satom9}.
The bosonic field content of the supergravity is the same 
as that of the usual (massless) 11D
supergravity:
the metric $\hat{g}_{\hat{\mu}\hat{\nu}}$  
and the 3-form gauge potential
$\hat{C}_{\hat{\mu}\hat{\nu}\hat{\rho}}$.
In this theory these fields are required to have a Killing isometry,
i.e., 
${\cal L}_{\hat{k}}\hat{g}_{\hat{\mu}\hat{\nu}}
={\cal L}_{\hat{k}}\hat{C}_{\hat{\mu}\hat{\nu}\hat{\rho}}=0$
where $ {\cal L}_{\hat{k}}$ indicates a Lie derivative 
with respect to a Killing vector field $\hat{k}^{\hat{\mu}}$.
(The coordinates are so chosen that 
the isometry direction is parametrized by the coordinate
$\hat{x}^{z}=z $, i.e.
$\hat{k}^{\hat{\mu}}
=\hat{\delta}^{\hat{\mu} z}$.)
The infinitesimal gauge transformations of the fields are
defined as\cite{berg4}
\beqa
\delta\hat{g}_{\hat{\mu}\hat{\nu}}&=&
-m[\hat{\lambda}_{\hat{\mu}}(i_{\hat{k}}\hat{g})_{\hat{\nu}}
+\hat{\lambda}_{\hat{\nu}}(i_{\hat{k}}\hat{g})_{\hat{\mu}}], 
\label{massivegt0}\\
\delta\hat{C}_{\hat{\mu}\hat{\nu}\hat{\rho}}&=&
3\hat{\partial}_{[\hat{\mu}}\hat{\chi}^{(2)}_{\hat{\nu}\hat{\rho}]}
-3m\hat{\lambda}_{[\hat{\mu}}(i_{\hat{k}}\hat{C})_{\hat{\nu}\hat{\rho}]}
\label{massivegt1}
\eeqa
where 
$(i_{\hat{k}}\hat{T}^{(r)}_{\hat{\mu}_{1}\cdots \hat{\mu}_{r-1}})
\equiv\hat{k}^{\hat{\mu}}
\hat{T}^{(r)}_{\hat{\mu}_{1}\cdots \hat{\mu}_{r-1}\hat{\mu}}$ 
for a field $\hat{T}^{(r)}$.
$\hat{\chi}^{(2)}_{\hat{\mu}\hat{\nu}}$ 
is the infinitesimal 2-form gauge parameter,
and $\hat{\lambda}_{\hat{\mu}}$ is defined as
$\hat{\lambda}_{\hat{\mu}}
\equiv (i_{\hat{k}}\hat{\chi}^{(2)}
)_{\hat{\mu}}$.\footnote{In this paper
we change the notation
of ref.\cite{berg4}
such that $m\to 2m$ and $\hat{\lambda}\to -\frac{1}{2}\hat{\lambda}$.
}
We note that the transformation corresponding to $\hat{\lambda}$ 
is called ``massive gauge transformation''.
Then, a connection for the massive gauge transformations  
must be introduced.
The new total connection takes the form 
$\hat{\Omega}_{\hat{a}}^{\ \hat{b}\hat{c}}
=\hat{\omega}_{\hat{a}}^{\ \hat{b}\hat{c}}
+\hat{K}_{\hat{a}}^{\ \hat{b}\hat{c}}$ 
where $\hat{\omega}_{\hat{a}}^{\ \hat{b}\hat{c}}$ 
is a usual spin connection
and
$\hat{K}$ is given by\cite{berg4}
\beqa
\hat{K}_{\hat{a}}^{\ \hat{b}\hat{c}}=\frac{m}{2}
[\hat{k}_{\hat{a}}(i_{\hat{k}}\hat{C})^{\hat{b}\hat{c}}
+\hat{k}^{\hat{b}}(i_{\hat{k}}\hat{C})_{\hat{a}}^{\ \hat{c}}
-\hat{k}^{\hat{c}}(i_{\hat{k}}\hat{C})_{\hat{a}}^{\ \hat{b}}].
\eeqa
The 4-form field strength $\hat{G}^{(4)}$ of $ \hat{C}$
is defined as\cite{berg4} 
\beqa
\hat{G}^{(4)}_{\hat{\mu}\hat{\nu}\hat{\rho}\hat{\sigma}}
=4\hat{D}_{[\hat{\mu}}\hat{C}_{\hat{\nu}\hat{\rho}\hat{\sigma}]}
\equiv 
4\hat{\partial}_{[\hat{\mu}}\hat{C}_{\hat{\nu}\hat{\rho}\hat{\sigma}]}
+3m(i_{\hat{k}}\hat{C})_{[\hat{\mu}\hat{\nu}}
(i_{\hat{k}}\hat{C})_{\hat{\rho}\hat{\sigma}]}
\eeqa
where $\hat{D}_{\hat{\mu}}$ denotes the covariant derivative.
Then, $\hat{G}^{(4)}$ transforms covariantly as
\beqa
\delta \hat{G}^{(4)}_{\hat{\mu}\hat{\nu}\hat{\rho}\hat{\sigma}}
=4m\hat{\lambda}_{[\hat{\mu}}
(i_{\hat{k}}\hat{G}^{(4)})_{\hat{\nu}\hat{\rho}\hat{\sigma}]},
\eeqa
which implies that $\delta (\hat{G}^{(4)})^{2}=0$.

The action of the massive 11D supergravity is\cite{berg4}
\beqa
\hat{S}_{0}&=&\frac{1}{\hat{\kappa}}
\int d^{11}\hat{x} [\ \sqrt{|\hat{g}|}\{ \hat{R}
-\frac{1}{2\cdot 4!}(\hat{G}^{(4)})^{2}
+\frac{1}{2}m^{2}|\hat{k}|^{4} \} \nonumber\\
&+&\frac{\hat{\epsilon}^{\hat{\mu}_{1}\cdots \hat{\mu}_{11}}}
{(144)^{2}}
\{2^{4}\hat{\partial}\hat{C}\hat{\partial}\hat{C}\hat{C}
+18m\hat{\partial}\hat{C}\hat{C}(i_{\hat{k}}\hat{C})^{2}
+\frac{3^{3}}{5}m^{2}\hat{C}(i_{\hat{k}}
\hat{C})^{4}\}_{
\hat{\mu}_{1}\cdots \hat{\mu}_{11}}]\ \ \ \ 
\label{11daction0}
\eeqa
where $\hat{\kappa} =16\pi G_{N}^{(11)}$ and $|\hat{k}|
=\sqrt{-\hat{k}^{\hat{\mu}}\hat{k}^{\hat{\nu}}
\hat{g}_{\hat{\mu}\hat{\nu}}}$.\footnote{
By using the (generalized) Palatini's identity
given in ref.\cite{berg4},
the action (\ref{11daction0}) is rewritten as
\beqa
\hat{S}_{0}&=&\frac{1}{\hat{\kappa}}
\int d^{11}\hat{x} [\sqrt{|\hat{g}|}
\{- \hat{\Omega}_{\hat{b}}^{\ \hat{b}\hat{a}}
\hat{\Omega}_{\hat{c}\ \hat{a}}^{\ \hat{c}}
-\hat{\Omega}_{\hat{a}}^{\ \hat{b}\hat{c}}
\hat{\Omega}_{\hat{b}\hat{c}}^{\ \ \hat{a}}
-\frac{1}{2\cdot 4!}(\hat{G}^{(4)})^{2}
+\frac{1}{2}m^{2}|\hat{k}^{2}|^{2} \} \}\nonumber\\
&+&\frac{1}{144}\hat{\epsilon}^{\hat{\mu}_{1}\cdots \hat{\mu}_{10} z}
\{ \hat{\partial}\hat{C}\hat{\partial}\hat{C}(i_{\hat{k}}\hat{C})
+\frac{m}{2}\hat{\partial}\hat{C}(i_{\hat{k}}\hat{C})^{3}
+\frac{9m^{2}}{80}(i_{\hat{k}}\hat{C})^{5}
\}_{\hat{\mu}_{1}\cdots\hat{\mu}_{10} z}
+({\rm surface\ \  terms})]. \nonumber
\eeqa
In fact, the 10D massive IIA action in ref.\cite{berg3} 
is obtained from this action
only if the surface terms are omitted.
Omitting them, we use this action as
a ``starting'' action, in order to 
make the correspondence of the 11D theory with the 10D one.}
This action is invariant up to total derivative
under the gauge transformation (\ref{massivegt0})
and (\ref{massivegt1}).


The 10-form gauge potential $\hat{A}^{(10)}$ to which
the M-9-brane minimally couples
is introduced\cite{satom9} by 
promoting the mass parameter $m$ to a scalar field $\hat{M}(\hat{x})$,
and adding to the action $\hat{S}_{0}$ the extra term 
\beqa
\Delta \hat{S} =\frac{1}{\hat{\kappa}}\int d^{11}\hat{x} 
\frac{1}{11!}\hat{\epsilon}^{\hat{\mu}_{1}\cdots\hat{\mu}_{11}}
\hat{M}(\hat{x})
11\hat{\partial}_{[\hat{\mu}_{1}} 
\hat{A}^{(10)}_{\hat{\mu}_{2}\cdots\hat{\mu}_{11}]}.
\label{deltaA}  
\eeqa
At this moment, the gauge transformation of the original action 
$\hat{S}_{0}$ does not vanish but is proportional to 
$\hat{\partial}\hat{M}$.
So, the total action $\hat{S}^{{\rm total}}\equiv \hat{S}_{0}
+\Delta  \hat{S}$
becomes invariant under (\ref{massivegt0}) and (\ref{massivegt1})
if the massive gauge transformation of $\hat{A}^{(10)}$
is defined as\cite{satom9}
\beqa
\delta (i_{\hat{k}}\hat{A}^{(10)})_{\hat{\mu}_{1}\cdots\hat{\mu}_{9}}
&=&-\sqrt{|\hat{g}|}
\hat{\epsilon}_{\hat{\mu}_{1}\cdots\hat{\mu}_{9}\hat{\mu} z}
[ 
-\hat{g}^{\hat{\mu}\hat{\mu}'}\hat{g}^{\hat{\nu}\hat{\nu}'}
(2\hat{\partial}_{[\hat{\mu}'} \hat{k}_{\hat{\nu}']}
-\hat{M}|\hat{k}|^{2}(i_{\hat{k}}\hat{C})_{\hat{\mu}'\hat{\nu}'})
\hat{\lambda}_{\hat{\nu}}\nonumber\\
&+&\frac{1}{2}\hat{G}^{(4)\hat{\mu}\hat{\nu}\hat{\rho}\hat{\sigma}}
(i_{\hat{k}}\hat{C})_{\hat{\nu}\hat{\rho}}\hat{\lambda}_{\hat{\sigma}}
]
-\frac{9!}{48}[\hat{\partial} 
\hat{C}(i_{\hat{k}}\hat{C})^{2}\hat{\lambda}
+\frac{\hat{M}}{4}(i_{\hat{k}}\hat{C})^{4}
\hat{\lambda}]_{\hat{\mu}_{1}\cdots\hat{\mu}_{9}}.\label{10formtr2}
\eeqa
We note that the 10-form $\hat{A}^{(10)}$ with {\it no} $z$ index
does not arise in the theory because $\hat{A}^{(10)}$ enters the
theory through the additional action
(\ref{deltaA}) and that $\hat{A}^{(10)}$ satisfies
${\cal L}_{\hat{k}}\hat{A}^{(10)}=0$.
We also note that the 10-form is not a dynamical field
(even if $\hat{M}$ is integrated out and the term like its kinetic
term arises in the action)
in the same way as the case of the RR 9-form potential
in 10 dimensions (see ref.\cite{pol2}).

In order to construct the gauge invariant M-9-brane WZ action
(and to derive an appropriate expression of field strength of
$\hat{A}^{(10)}$), 
dual fields of the 3-form $\hat{C}$ and a ``1-form potential''
$\hat{k}_{\hat{\mu}}$ need to be 
introduced\cite{satom9}.\footnote{The problem was that 
the first part of (\ref{10formtr2}) cannot be expressed as a sum of
products of forms, but we can rewrite the part into that
if dual fields are used appropriately through duality relations.}
They are the 6-form and the 8-form potentials
$\hat{C}^{(6)}$ and $\hat{N}^{(8)}$, to which the M-5-brane and
M-KK-monopole minimally couple\cite{berg4}\cite{loz1}, 
respectively. 
Their gauge transformations for a constant mass background $\hat{M}=m$
are defined as\cite{berg4}
\beqa
\delta \hat{C}^{(6)}_{\hat{\mu_{1}}\cdots\hat{\mu_{6}}}&=&
6\hat{\partial}_{[\hat{\mu}_{1}}
\hat{\chi}^{(5)}_{\hat{\mu}_{2}\cdots\hat{\mu}_{6}]}
+30\hat{\partial}_{[\hat{\mu}_{1}}
\hat{\chi}^{(2)}_{\hat{\mu}_{2}\hat{\mu}_{3}}
\hat{C}_{\hat{\mu}_{4}\hat{\mu}_{5}\hat{\mu}_{6}]}  
+6 m \hat\lambda_{[\hat{\mu}_{1}}
(i_{\hat{k}}\hat{C}^{(6)})_{\hat{\mu}_{2}\cdots\hat{\mu}_{6}]}\\
\delta \hat{N}^{(8)}_{\hat{\mu}_{1}\cdots\hat{\mu}_{8}}&=&\{
8\hat{\partial}\hat{\Omega}^{(7)}
+168\hat{\partial}\hat{\chi}^{(5)}(i_{\hat{k}}\hat{C})
+\frac{8!}{3\cdot 4!}\hat{\partial}\hat{\chi}^{(2)}\hat{C}
(i_{\hat{k}}\hat{C})
+8 m \hat{\lambda}(i_{\hat{k}}\hat{N}^{(8)})
\}_{[\hat{\mu}_{1}\cdots\hat{\mu}_{8}]}\ \ \ \ \ \ \ \ \ 
\eeqa
where $\hat{\chi}^{(5)}$ and $\hat{\Omega}^{(7)}$ are 5-form and
7-form parameters of the gauge transformation associated with 
$\hat{C}^{(6)}$ and $\hat{N}^{(8)}$, respectively.
The duality relations which they satisfy are\cite{berg4}\cite{satom9}
\beqa
\hat{G}^{(4)\hat{\mu}_{1}\cdots\hat{\mu}_{4}}
=\frac{\hat{\epsilon}^{\hat{\mu}_{1}\cdots
\hat{\mu}_{11}}}{7!\sqrt{|\hat{g}|}}\hat{G}^{(7)}_{\hat{\mu}_{5}\cdots
\hat{\mu}_{11}},\ \ \hat{G}^{(2)\hat{\mu}_{1}\hat{\mu}_{2}}
=\frac{\hat{\epsilon}^{\hat{\mu}_{1}\cdots
\hat{\mu}_{10}z}}{9!\sqrt{|\hat{g}|}}
(i_{\hat{k}}\hat{G}^{(9)})_{\hat{\mu}_{3}\cdots\hat{\mu}_{10}}
\label{duality2}
\eeqa
where 
\beqa
\hat{G}^{(4)}_{\hat{\mu}_{1}\cdots\hat{\mu}_{4}}&\equiv&
\{4\hat{\partial} \hat{C}
+3m(i_{\hat{k}}\hat{C})(i_{\hat{k}}\hat{C})
\}_{\hat{\mu}_{1}\cdots\hat{\mu}_{4}}\\
\hat{G}^{(7)}_{\hat{\mu}_{1}\cdots\hat{\mu}_{7}}&\equiv&7\{
\hat{\partial}\hat{C}^{(6)}
-3 m (i_{\hat{k}}\hat{C})(i_{\hat{k}}\hat{C}^{(6)})
+10\hat{C}\hat{\partial} \hat{C}\nonumber\\
& &+5 m \hat{C}(i_{\hat{k}}\hat{C})^{2}
+\frac{m}{7}(i_{\hat{k}}\hat{N}^{(8)})
\}_{\hat{\mu}_{1}\cdots\hat{\mu}_{7}}\\
\hat{G}^{(2)}_{\hat{\mu}\hat{\nu}}&\equiv& 
2\hat{\partial}_{[\hat{\mu}}\hat{k}_{\hat{\nu}]}
-m |\hat{k}|^{2}(i_{\hat{k}}\hat{C})_{\hat{\mu}\hat{\nu}}\\
(i_{\hat{k}}\hat{G}^{(9)})_{\hat{\mu}_{1}\cdots\hat{\mu}_{8}}&\equiv&
8\{ \hat{\partial} (i_{\hat{k}}\hat{N}^{(8)})
+21(i_{\hat{k}}\hat{C}^{(6)})\hat{\partial}(i_{\hat{k}}\hat{C})
\nonumber\\
&+&35\hat{C}\hat{\partial}(i_{\hat{k}}\hat{C})(i_{\hat{k}}\hat{C})
+35\hat{\partial} \hat{C}(i_{\hat{k}}\hat{C})^{2}
+\frac{105}{8} m (i_{\hat{k}}\hat{C})^{4}
\}_{[\hat{\mu}_{1}\cdots\hat{\mu}_{8}]}\label{duality1}
\eeqa
are the field strengths of $\hat{C},\hat{C}^{(6)},\hat{k}_{\mu}$ and
$i_{\hat{k}}\hat{N}^{(8)}$, respectively.\footnote{
In the case of $\hat{N}^{(8)}$, its full field strength
is difficult to construct, but that of $i_{\hat{k}}\hat{N}^{(8)}$ 
can be obtained, and it is sufficient for the present 
purpose\cite{satom9}.}

By using (\ref{duality2}), the gauge invariant WZ action of the
M-9-brane is constructed\cite{satom9}. 
The field strength of $\hat{A}^{(10)}$ can also be defined, but
in order to discuss dimensional reductions, it is convenient to 
introduce a 10-form potential $\hat{C}^{(10)}$ which gives
usual RR 9-form potential $C^{(9)}$ in the massive IIA theory 
on dimensional reduction along $z$.
Based on the transformation property under the massive transformation,
$\hat{C}^{(10)}$ is identified as the following redefined field:
\beqa
(i_{\hat{k}}\hat{C}^{(10)})_{\hat{\mu}_{1}\cdots\hat{\mu}_{9}}
&\equiv&(i_{\hat{k}}\hat{A}^{(10)})_{\hat{\mu}_{1}\cdots\hat{\mu}_{9}}
+9!\ [\frac{1}{2\cdot
7!}(i_{\hat{k}}\hat{N}^{(8)})(i_{\hat{k}}\hat{C})
\nonumber\\
& &-\frac{1}{2^{3}\cdot 5!}(i_{\hat{k}}\hat{C}^{(6)})
(i_{\hat{k}}\hat{C})^{2}
+\frac{1}{2^{4}\cdot (3!)^{2}}
\hat{C}(i_{\hat{k}}\hat{C})^{3}]_{[\hat{\mu}_{1}\cdots\hat{\mu}_{9}]}.
\eeqa
The 11-form field strength $\hat{G}^{(11)}$
of $\hat{C}^{(10)}$ is given as the equation of motion for $\hat{M}$,
as
\beqa
\frac{\delta\hat{S}^{total}}{\delta \hat{M}}=0 \ \Leftrightarrow \ \ 
\hat{M}|\hat{k}|^{4}=-*\hat{G}^{(11)}\label{11dmassdual}
\eeqa
where $*$ means Hodge or Poincare dual, and
\beqa
\hat{G}^{(11)}
=(i_{\hat{k}}\hat{G}^{(11)})_{\hat{\mu}_{1}\cdots\hat{\mu}_{10}}&=&
10 \{ \hat{\partial}(i_{\hat{k}}\hat{C}^{(10)})
-36\hat{\partial}
(i_{\hat{k}}\hat{C})(i_{\hat{k}}\hat{N}^{(8)})\nonumber\\
& &-36\cdot 35
\hat{C}\hat{\partial}(i_{\hat{k}}\hat{C})(i_{\hat{k}}\hat{C})^{2}
+\frac{189}{2}
m(i_{\hat{k}}\hat{C})^{5}\}_{\hat{\mu}_{1}\cdots\hat{\mu}_{10}}.
\eeqa
Then, the gauge transformations of $i_{\hat{k}}\hat{C}^{(10)}$
can be defined so as to keep $\hat{G}^{(11)}$ invariant\cite{satom9}, 
as
\beqa
\delta (i_{\hat{k}}\hat{C}^{(10)})_{\hat{\mu}_{1}\cdots\hat{\mu}_{9}}
&=&9!\{
\frac{1}{8!}\hat{\partial} (i_{\hat{k}}\hat{\Omega}^{(9)})
+\frac{1}{2\cdot 6!}\hat{\partial}(i_{\hat{k}}\hat{\Omega}^{(7)})
(i_{\hat{k}}\hat{C})
+\frac{1}{2^{3}\cdot 4!}\hat{\partial}(i_{\hat{k}}\hat{\chi}^{(5)})
(i_{\hat{k}}\hat{C})^{2}\nonumber\\
&+&\frac{1}{2^{4}\cdot 3!}
\hat{\partial}\hat{\chi}^{(2)}(i_{\hat{k}}\hat{C})^{3}
-\frac{1}{2^{4}\cdot 4!}m\hat{\lambda}(i_{\hat{k}}\hat{C})^{4}
\}_{[\hat{\mu}_{1}\cdots\hat{\mu}_{9}]},
\label{c10gaugetr}
\eeqa 
where $\hat{\Omega}^{(9)}$ is a 9-form parameter of the gauge
transformation associated with $\hat{C}^{(10)}$.



\subsection{Review of the M-9-brane WZ action}
In this approach, 
the M-9-brane
wrapped around the compact isometry direction is 
described\cite{bergm9} (see also \cite{hulm9}), and
its worldvolume effective action is constructed as that of a
gauged $\sigma$-model\cite{bergm9}\cite{loz1}\cite{satom9}, 
where the the translation along
$\hat{k}^{\hat{\mu}}$ is 
gauged\cite{berg4} (see also ref.\cite{kaluzakleinm}).
Denoting its worldvolume coordinates by $\xi^{i} \  (i=0,1,..,8)$
and their embeddings by $\hat{X}^{\hat{\mu}}(\xi)$
($ \hat{\mu}=0,1,\cdots,9,z$),
the 
worldvolume gauge
transformation is written by
\beqa
\delta_{\eta} \hat{X}^{\hat{\mu}}
=\eta(\xi)\hat{k}^{\hat{\mu}},
\label{wvgaugetr}
\eeqa
where 
$\eta(\xi)$ is a scalar gauge parameter.
To make the brane action invariant
under the transformation,
the derivative of $\hat{X}^{\hat{\mu}}$ with respect to $\xi^{i}$ is 
replaced by the covariant derivative
$D_{i}\hat{X}^{\hat{\mu}}=\partial_{i}\hat{X}^{\hat{\mu}}
-\hat{A}_{i}\hat{k}^{\hat{\mu}} $
with the gauge field $\hat{A}_{i}=-|\hat{k}|^{-2}
\partial_{i}\hat{X}^{\hat{\nu}}\hat{k}_{\hat{\nu}}$\cite{kaluzakleinm}.
($D_{i}\hat{X}^{\hat{\mu}}$ is invariant under (\ref{wvgaugetr}).)
Then, {\it only on the basis of the invariance under the massive gauge
transformation} (and (\ref{wvgaugetr})),
the M-9-brane WZ action (for a constant mass background)
is constructed as\cite{satom9}
\beqa
S_{M9}^{WZ}\ \ \ =\ \ \ 
\int d^{9}\xi \hat{\epsilon}^{i_{1}\cdots i_{9}} 
[\frac{1}{9!}\widetilde{(i_{\hat{k}}\hat{C}^{(10)})}_{i_{1}\cdots i_{9}}
+\frac{1}{2\cdot 7!}
\widetilde{(i_{\hat{k}}\hat{N}^{(8)})}_{i_{1}\cdots i_{7}}
\hat{{\cal K}}^{(2)}_{i_{8}i_{9}}\ \ \ \ \ \ \ \ \ \ \ \ \ \ \ 
\ \ \ \ \ \ \  
\nonumber\\
+\frac{1}{2^{3}\cdot 5!}\widetilde{(i_{\hat{k}}\hat{C}^{(6)})}_{
i_{1}\cdots i_{5}}
(\hat{{\cal K}}^{(2)})^{2}_{{i_{6}\cdots i_{9}}} 
+\frac{1}{2\cdot (3!)^{2}}\widetilde{\hat{C}}_{i_{1} i_{2} i_{3}}
\hat{{\cal K}}^{(2)}_{i_{4} i_{5}}
\{(\partial \hat{b})^{2} 
-\frac{1}{4}\widetilde{(i_{\hat{k}}\hat{C})}
\partial \hat{b} 
+\frac{1}{8}\widetilde{(i_{\hat{k}}\hat{C})}^{2}\}_{i_{6}\cdots i_{9}}
\nonumber\\
+\frac{1}{2\cdot 4!}\hat{A}_{i_{1}}
\hat{{\cal K}}^{(2)}_{i_{2} i_{3}}
\{ (\partial \hat{b})^{3} 
+\frac{1}{2}(\partial \hat{b})^{2}\widetilde{(i_{\hat{k}}\hat{C})}
+\frac{1}{4}(\partial \hat{b})\widetilde{(i_{\hat{k}}\hat{C})}^{2}
+\frac{1}{8}\widetilde{(i_{\hat{k}}\hat{C})}^{3}
\}_{i_{4}\cdots i_{9}}\nonumber\\
+\frac{m}{5!}\hat{b}_{i_{1}}(\partial \hat{b})^{4}_{i_{2}\cdots
i_{9}}
+\frac{1}{8!}\partial_{i_{1}} 
\hat{\omega}^{(8)}_{i_{2}\cdots  i_{9}}]\ \ \ \ \ 
\label{m9action}
\eeqa
where $\widetilde{\hat{T}^{(r)}}_{i_{1}\cdots i_{r}}\equiv 
\hat{T}^{(r)}_{\hat{\mu}_{1}\cdots\hat{\mu}_{r}}
D_{i_{1}}\hat{X}^{\hat{\mu}_{1}}
\cdots D_{i_{r}}\hat{X}^{\hat{\mu}_{r}}$
for a target-space r-form field 
$\hat{T}^{(r)}_{\hat{\mu}_{1}\cdots\hat{\mu}_{r}}$.
$\hat{b}_{i}$ 
describes the flux 
of an M-2-brane wrapped around the isometry direction\cite{loz1},
whose massive gauge transformation is determined by the 
requirement of the invariance of its modified field strength
$\hat{{\cal K}}^{(2)}_{ij}=2\partial_{[i}\hat{b}_{j]}-
D_{i}\hat{X}^{\hat{\mu}}D_{j}\hat{X}^{\hat{\nu}}
(i_{\hat{k}}\hat{C})_{\hat{\mu}\hat{\nu}}$ (i.e. $\delta\hat{b}_{i}
=\hat{\lambda}_{i}$).
We note that $S_{M9}^{WZ}$ is exactly gauge invariant if the
worldvolume 8-form $\hat{\omega}^{(8)}$ is transformed as
\beqa
\delta\hat{\omega}^{(8)}=8!\{
\frac{1}{6!}\partial (i_{\hat{k}}\hat{\Omega}^{(7)}) \hat{b}
+\frac{1}{2\cdot 4!}
\partial(i_{\hat{k}}\hat{\chi}^{(5)}) \hat{b}\partial \hat{b}
+\frac{1}{2\cdot 3!}\partial \hat{\chi}^{(2)}
\hat{b}(\partial \hat{b})^{2}
-\frac{4}{5!}m\hat{\lambda}\hat{b}(\partial \hat{b})^{3}
\}.
\eeqa

Actually, 
the expression of the action using the covariant derivative $DX$
is not so convenient to discuss dimensional reductions of 
the action $S_{M9}^{WZ}$.
So, we rewrite the action into the expression without $DX$
using the two identities:
\beqa
\widetilde{(i_{\hat{k}}\hat{T}^{(r)})}
_{i_{1}\cdots i_{r-1}}=
(i_{\hat{k}}\hat{T}^{(r)})_{i_{1}\cdots i_{r-1}},\ \ 
\widetilde{\hat{T}^{(r)}}_{i_{1}\cdots
i_{r}}=\hat{T}^{(r)}_{i_{1}\cdots
i_{r}}-r\cdot \hat{A}_{[i_{1}}(i_{\hat{k}}\hat{T}^{(r)})
_{i_{2}\cdots i_{r}]},\label{identity}
\eeqa
where $\hat{T}^{(r)}_{i_{1}\cdots i_{r}}\equiv
\hat{T}^{(r)}_{\hat{\mu}_{1}\cdots \hat{\mu}_{r}}
\partial_{i_{1}}\hat{X}^{\hat{\mu}_{1}}\cdots \partial_{i_{1}}
\hat{X}^{\hat{\mu}_{1}}$.
We note that 
$D_{i}\hat{X}^{\hat{\mu}}=\partial_{i}\hat{X}^{\hat{\mu}}$ for
$\hat{\mu}\ne z$ and
$D_{i}\hat{X}^{z}=-\hat{A}_{i}$.
Then, the WZ action (\ref{m9action}) is written as the slightly simple
form:
\beqa
S_{M9}^{WZ}\ \ \ =\ \ \ 
\int d^{9}\xi \hat{\epsilon}^{i_{1}\cdots i_{9}} 
[\frac{1}{9!}(i_{\hat{k}}\hat{C}^{(10)})_{i_{1}\cdots i_{9}}
+\frac{1}{2\cdot 7!}
(i_{\hat{k}}\hat{N}^{(8)})_{i_{1}\cdots i_{7}}
\hat{{\cal K}}^{'(2)}_{i_{8}i_{9}}\ \ \ \ \ \ \ \ \ \ \ \ \ \ \ 
\ \ \ \ \ \ \  
\nonumber\\
+\frac{1}{2^{3}\cdot 5!}(i_{\hat{k}}\hat{C}^{(6)})_{
i_{1}\cdots i_{5}}
(\hat{{\cal K}}^{'(2)})^{2}_{{i_{6}\cdots i_{9}}} 
+\frac{1}{2\cdot (3!)^{2}}\hat{C}_{i_{1} i_{2} i_{3}}
\hat{{\cal K}}^{'(2)}_{i_{4} i_{5}}
\{(\partial \hat{b})^{2} 
-\frac{1}{4}(i_{\hat{k}}\hat{C})
\partial \hat{b} 
+\frac{1}{8}(i_{\hat{k}}\hat{C})^{2}\}_{i_{6}\cdots i_{9}}
\nonumber\\
+\frac{1}{2^{4}\cdot 4!}\hat{A}_{i_{1}}
(\hat{{\cal K}}^{'(2)})^{4}_{i_{2}\cdots i_{9}}
+\frac{m}{5!}\hat{b}_{i_{1}}(\partial \hat{b})^{4}_{i_{2}\cdots
i_{9}}
+\frac{1}{8!}\partial_{i_{1}} 
\hat{\omega}^{(8)}_{i_{2}\cdots  i_{9}}]\ \ \ \ \ 
\label{m9action2}
\eeqa
where $\hat{{\cal K}}^{'(2)}_{ij}=2\partial_{[i}\hat{b}_{j]}-
\partial_{i}\hat{X}^{\hat{\mu}}\partial_{j}\hat{X}^{\hat{\nu}}
(i_{\hat{k}}\hat{C})_{\hat{\mu}\hat{\nu}}$.
We use this form to discuss dimensional reductions of
$S_{M9}^{WZ}$.

\section{Dimensional reductions of the M-9-brane WZ action}
\setcounter{footnote}{0} 

Now we discuss the dimensional reductions of $S_{M9}^{WZ}$.
Since the purpose of this work is to examine the consistency,
the important point is that {\it precisely the same forms of the actions 
are derived}. Thus, in each case of the reductions, we present
explicit correspondence of the terms in $S_{M9}^{WZ}$
with those in the WZ action of the other brane.

\subsection{The dimensional reduction along the isometry direction}

First, we discuss the dimensional reduction of $S_{M9}^{WZ}$
along the isometry direction
and show that WZ action of
the D-8-brane {\it with no isometry direction} is derived.
In this case we split the coordinates $\hat{x}^{\hat{\mu}}$ for 
$\hat{\mu}=0,1,\cdots,9,z$ into
$(x^{\mu},z)$ ($\mu=0,1,\cdots,9$). 
Since one of the worldvolume directions of the M-9-brane 
is considered to be wrapped around
the isometry direction,
all we have to do is to rewrite $S_{M9}^{WZ}$ in terms of  
10D fields.

The relations of original 11D target-space
fields to the 10D ones can
be taken as the familiar form\cite{berg4}
\beqa
\left. 
\begin{array}{cc}
\left\{
\begin{array}{cc}
\hat{g}_{\mu\nu}= &e^{-2\phi/3}g_{\mu\nu}
- e^{4\phi/3}C^{(1)}_{\mu}C^{(1)}_{\nu}, \\
\hat{g}_{\mu z}= & (i_{\hat{k}}\hat{g})_{\mu}=
- e^{4\phi/3}C^{(1)}_{\mu}  \\
\hat{g}_{z z}= &-e^{4\phi/3}
\end{array}
\right. 
&
\left\{
\begin{array}{cc}
\hat{C}_{\mu\nu\rho}= &C^{(3)}_{\mu\nu\rho} \\
\hat{C}_{\mu\nu z}= &(i_{\hat{k}}\hat{C})_{\mu\nu}=B_{\mu\nu}.
\end{array}
\right.
\end{array}
\right.\label{relofgc}
\eeqa
where $B$ is the NSNS 2-form 
and $C^{(1)}$ and $C^{(3)}$ 
are the 10D RR 1-form and 3-form 
potentials (we denote 10D RR r-forms as $C^{(r)}$).
By using these relations, 
the massive 11D action (\ref{11daction0}) 
is shown to give the bosonic part 
of the action of 10D massive IIA supergravity 
given in ref.\cite{berg3} on the dimensional reduction\cite{berg4}.
In particular, the third term of (\ref{11daction0})
gives rise to the cosmological term $-m^{2}/2$, and
the kinetic term of $\hat{C}$ gives a mass term of $B_{\mu\nu}$.

The dimensional reductions of the other target-space
gauge fields are given 
as\cite{berg4}\cite{satom9}
\beqa
\left\{
\begin{array}{cc}
\hat{C}^{(6)}_{\mu_{1}\cdots\mu_{6}}= &
-\tilde{B}^{(6)}_{\mu_{1}\cdots\mu_{6}}, \\
\hat{C}^{(6)}_{\mu_{1}\cdots\mu_{5} z}= &
(i_{\hat{k}}\hat{C}^{(6)})_{\mu_{1}\cdots\mu_{5}}=
C^{(5)}_{\mu_{1}\cdots\mu_{5}}-5C^{(3)}_{[\mu_{1}\mu_{2}\mu_{3}}
B_{\mu_{4}\mu_{5}]},\label{relofc6}
\end{array}
\right.
\eeqa
\beqa
(i_{\hat{k}}\hat{N}^{(8)})_{\mu_{1}\cdots\mu_{7}}&=&
C^{(7)}_{\mu_{1}\cdots\mu_{7}}-7\cdot 5C^{(3)}_{[\mu_{1}\mu_{2}\mu_{3}}
B_{\mu_{4}\mu_{5}}B_{\mu_{6}\mu_{7}]}\label{relofn8}\\
(i_{\hat{k}}\hat{C}^{(10)})_{\mu_{1}\cdots\mu_{9}}
&=&C^{(9)}_{\mu_{1}\cdots\mu_{9}}\label{relofc10}
\eeqa
where $\tilde{B}^{(6)}$ is 
the NSNS 6-form gauge field, the dual of $B$. 
The relations of worldvolume fields are that $\hat{b}_{i}$ corresponds
to the BI field on the D-8-brane $V_{i}$, and 
$\hat{\omega}^{(8)}$ corresponds to an 8-form $\omega^{(8)}$. 

Then, dimensionally reducing $S_{M9}^{WZ}$ along $z$, 
we have the following action 
\beqa
S^{'WZ}=\int  d^{9}\xi \epsilon^{i_{1}\cdots i_{9}}[
\sum_{r=1,3,5,7,9}\{ \frac{1}{r!q!2^{q}}C^{(r)}{\cal F}^{q} \}
+\frac{m}{5!}V(\partial V)^{4}
+\frac{1}{8!}\partial 
\omega^{(8)}]_{i_{1}\cdots  i_{9}}\label{d8action}
\eeqa
where ${\cal F}_{ij}\equiv 2\partial_{[i}V_{j]}-B_{ij}$, 
$q=(9-r)/2$ and $T^{(r)}_{i_{1}\cdots i_{r}}\equiv 
T^{(r)}_{\mu_{1}\cdots \mu_{r}}\partial_{i_{1}}X^{\mu_{1}}
\cdots \partial_{i_{r}}X^{\mu_{r}}$ for a field 
$T^{(r)}_{\mu_{1}\cdots \mu_{r}}$.
This is exactly the bosonic part of the D-8-brane WZ action
in a massive IIA background\cite{massDbr2}\cite{massDbr3}.
Thus, we can say that
{\it this reduction of the M-9-brane is consistent with 
the relations of the branes given in Fig.1.}

The explicit correspondence of the terms in $S_{M9}^{WZ}$ with those
in $S_{D8}^{WZ}$ is given as follows:
By using (\ref{relofc6}), (\ref{relofn8}) and (\ref{relofc10}),
the first three terms of (\ref{m9action2}) give the terms
including $C^{(9)}$,  $C^{(7)}$ and $C^{(5)}$
in the action (\ref{d8action}), 
in addition to the contribution
\beqa
\frac{1}{2^{3}(3!)^{2}}\{C^{(3)}{\cal F}
(\frac{B^{2}}{2}-3B\partial V)\}_{i_{1}\cdots i_{9}},
\label{cfbb}
\eeqa
since $\hat{{\cal K}}^{'(2)}_{ij}$ gives ${\cal F}_{ij}$.
This term (\ref{cfbb}), 
combined with the contribution of the forth term in 
(\ref{m9action2}) gives the term including $C^{(3)}$ 
in the action (\ref{d8action}).
Finally, the fifth term of (\ref{m9action2}) 
(including $\hat{A}$) gives
the term including $C^{(1)}$ 
in the action (\ref{d8action}) since it holds $\hat{A}_{i}=
C^{(1)}_{\mu}\partial_{i}X^{\mu}$.


\subsection{The dimensional reduction along a standard
worldvolume direction}

Next, we show that on this double dimensional reduction,
the action $S^{''WZ}$ obtained from $S_{M9}^{WZ}$ can be identified
with the WZ action of the KK-8A brane $S_{KK8A}^{WZ}$ 
presented in ref.\cite{exotic}, which we consider to be  
a D-8-brane with an isometry direction due to some special
background.

The outline of this identification is as follows:
First, we derive the action $S^{''WZ}$ and 
discuss the difference of 
the obtained action $S^{''WZ}$ from $S^{WZ}_{KK8A}$. 
Then, 
the two looks certainly different from each other 
at a glance.
However, there is a target-space field appearing in each of  
the action whose relation to each other is not known yet;
a 9-form potential $C^{'(9)}$ appearing in $S^{''WZ}$ which
originates from the 11D 10-form 
$\hat{C}^{(10)}$, and a 9-form $N^{(9)}$
appearing in $S_{KK8A}^{WZ}$\cite{exotic} which
comes from the 10D IIB 8-form via T-duality.  
The relation of these two 9-forms has not been discussed,
while the other fields we use are completely 
the same as those in ref.\cite{exotic}.\footnote{
The distinctions of the fields are
made on the basis of their gauge transformation properties and
the 11D origins of the fields.}
Thus, we can say that {\it the two actions are equivalent
if the difference of the two can be cancelled
only by setting the field redefinition relation of the two 9-forms.}
In addition, we can check the consistency of the redefinition relation 
by discussing the gauge transformations of both sides of the
relation.
We will demonstrate these in the following.

In this case we denote the worldvolume
indices of the M-9-brane by $\hat{i}$, 
and split the coordinates 
$\hat{x}^{\hat{\mu}}$ 
into $(x^{\mu},x^{8})$ ($\mu=0,1,\cdots,7,9,z$) and 
$\xi^{\hat{i}} $ 
into $(\xi^{i},\xi^{8})$ 
($i=0,1,\cdots,7$). Then, we fix the coordinates
so that $X^{8}(\xi)=\xi^{8}$ and consider the double
dimensional reduction along $X^{8}=\xi^{8}$.
Please remember 
that on this reduction,
one of the indices $i_{r}$ ($r=1,\cdots,9$)
in the M-9-brane WZ action 
(\ref{m9action2})
always takes the value of 8,
and that it holds 
$\partial_{8}\hat{X}^{\hat{\mu}}=\delta^{\hat{\mu}}_{8} $.


As for the target-space fields,
the relations of $\hat{g}_{\hat{\mu}\hat{\nu}}$ and $\hat{C}$ 
with the 10D fields are the same as (\ref{relofgc})
except for the replacement of the indices $z\to 8$,
but the obtained theory is not the usual massive IIA SUGRA in 
ref.\cite{rom1}\cite{berg3} 
because there remains an isometry direction. 
That is, 
this is another massive extension of standard (massless) IIA SUGRA.
However, 
based on the analyses on dimensional reductions of the
massive 11D SUGRA to 9-dimensional SUGRAs,
the authors of ref.\cite{meessen1} argue that
the obtained another massive IIA SUGRA is related to the usual massive
IIA one by a rotation in internal space.
We believe their argument and consider the field content 
appearing in this massive SUGRA to be essentially
the same as that in the usual massive IIA one.
We note that the 10D Killing vector $k^{\mu}$
is defined as $k^{\mu}\equiv \hat{k}^{\mu}$ ($\hat{k}^{8}=0$).


The dimensional reductions of the dual fields are as follows:
That of $\hat{C}^{(6)}$ is again the same as (\ref{relofc6})
except for the replacement $z\to 8$.
On the other hand,
the dimensional reduction of $i_{\hat{k}}\hat{N}^{(8)}$ is given 
by\cite{loz1}\cite{exotic}
\beqa
\left\{
\begin{array}{cc}
(i_{\hat{k}}\hat{N}^{(8)})_{\mu_{1}\cdots\mu_{7}}= &
-(i_{k}N^{(8)})_{\mu_{1}\cdots\mu_{7}}, \\
(i_{\hat{k}}\hat{N}^{(8)})_{\mu_{1}\cdots\mu_{6} y}= & 
(i_{k}N^{(7)})_{\mu_{1}\cdots\mu_{6}}
\end{array}
\right.
\eeqa
where 
$i_{k}N^{(8)}$ is the 8-form dual of the ``scalar field'' 
$(i_{k}C^{(1)})$ while $i_{k}N^{(7)}$ is the 7-form
dual of the 1-form field $k_{\mu}=(i_{k}g)_{\mu}$.
The duality relations they satisfy are derived from the second
equation of (\ref{duality2}).
The brane which couples to $i_{k}N^{(8)}$ is called
``KK-6A brane''\cite{meessen1}\cite{eyras1}\cite{exotic},
the solution corresponding to which is shown to be identified with
the D-6-brane solution via a coordinate transformation\cite{exotic}.
The brane which couples to 
$i_{k}N^{(7)}$ is a IIA KK-monopole.
We note that $\hat{N}^{(8)}$ gives $N^{(7)}$ but not  
$ C^{(7)}$ unlike (\ref{relofn8}).
This is possible
because the definition of $i_{\hat{k}}\hat{N}^{(8)}$ itself,
the 11D origin of $N^{(7)}$, depends on $\hat{k}$.
(see eq.(\ref{duality2})).

The dimensional reduction of $(i_{\hat{k}}\hat{C}^{(10)})$ is 
written as
\beqa
\left\{
\begin{array}{cc}
(i_{\hat{k}}\hat{C}^{(10)})_{\mu_{1}\cdots\mu_{9}}= &
-(i_{k}B^{(10)})_{\mu_{1}\cdots\mu_{9}} \\
(i_{\hat{k}}\hat{C}^{(10)})_{\mu_{1}\cdots\mu_{8} y}= & 
-(i_{k}C^{'(9)})_{\mu_{1}\cdots\mu_{8}}
\end{array}
\right.
\eeqa
where $B^{(10)}$ is a NSNS 10-form to which the NS-9A brane 
couples\cite{sptfilling}. 
($B^{(10)}$ does not appear in $S_{KK8A}^{WZ}$
since in this reduction one of the indices of 
$i_{\hat{k}}\hat{C}^{(10)}$ arising in the action (\ref{m9action2})
takes the value of 8.)
$C^{'(9)}$ is considered to be a kind of RR 9-form potential, 
to which the KK-8A brane is expected to couple minimally.
The difference between this 9-form
$C^{'(9)}$ and the usual $C^{(9)}$ coming from the existence of 
the isometry appears in
the duality relations they satisfy. For a constant mass
background $\hat{M}=m$, they are given respectively as
\beqa
m|k|^{4}e^{-4\phi}&=& *(dC^{'(9)}+\cdots)\label{dualityofc9'}\\
m&=& *(dC^{(9)}+\cdots)\label{dualityofc9}
\eeqa
where $\cdots$ are the parts not essential to our discussion.
The gauge transformations of $B^{(10)}$ and
$ C^{'(9)}$ 
are obtained from (\ref{c10gaugetr}). 
We note that since $\hat{C}^{(10)}$ has no 
dynamical degrees of freedom,
so do $B^{(10)}$ and $ C^{'(9)}$.

As for the worldvolume gauge fields,
we split $\hat{b}_{\hat{i}}$ 
into $(\omega^{(1)}_{i}, \omega^{(0)})$, as ref.\cite{exotic},
and define a worldvolume 7-form as
the dimensional reduction of $\hat{\omega}^{(8)}$: $\omega^{(7)}\equiv
\hat{\omega}^{(8)}_{i_{1}\cdots i_{7}8}$.
(We note that $\hat{\omega}^{(8)}$ with no index of 8 
does not appear in the action.)
By using (\ref{relofgc}),
the dimensional reduction of $\hat{A}_{\hat{i}}\  (=
-|\hat{k}|^{-2}\partial_{\hat{i}}\hat{X}^{\hat{\mu}}
\hat{k}_{\hat{\mu}})$ 
is written as
\beqa
\hat{A}_{i}&=&\{1+e^{2\phi}|k|^{-2}(i_{k}C^{(1)})^{2}\}^{-1}
\{A_{i}+e^{2\phi}|k|^{-2}(i_{k}C^{(1)})C^{(1)}_{i}\}\nonumber\\
&=&
A_{i}+\{1+e^{2\phi}|k|^{-2}(i_{k}C^{(1)})^{2}\}^{-1}
e^{2\phi}|k|^{-2}(i_{k}C^{(1)})D_{i}X^{\mu} C^{(1)}_{\mu}
\label{ai2}\\
\hat{A}_{8}&=&\{1+e^{2\phi}|k|^{-2}(i_{k}C^{(1)})^{2}\}^{-1}
e^{2\phi}|k|^{-2}(i_{k}C^{(1)})
\eeqa
where $A_{i}\equiv -|k|^{-2}\partial_{i}X^{\mu}k_{\mu}$
and $D_{i}X^{\mu} \equiv \partial_{i}X^{\mu}-A_{i}k^{\mu}$
is the worldvolume covariant derivative on the KK-8A brane\cite{loz1}. 
For later use, we 
define the ``field strengths'' of $\omega^{(1)}$ and $\omega^{(0)}$
as
\beqa
{\cal K'}^{(2)}_{ij}&\equiv&
\hat{{\cal K}}^{(2)}_{ij}
=2\partial_{[i} \omega^{(1)}_{j]}
-(i_{k}C^{(3)})_{ij}\\
{\cal K}^{(1)}_{i} &\equiv&
\hat{{\cal K}}^{(2)}_{i8}
=\partial_{i} \omega^{(0)}+(i_{k}B)_{i}.
\eeqa


Now, let us discuss the difference between $S^{WZ}_{KK8A}$
and the action 
$S^{''WZ}$ which will be obtained from $S^{WZ}_{M9}$.
Since the action $S^{''WZ}$ and $S^{WZ}_{KK8A}$ 
are both too complicated to deal with at once,
we divide the contribution of $S_{M9}^{WZ}$ into 4 parts ;
the part (I): terms including p-form potentials with $p>3$, 
the part (II): those including $C^{(3)}_{ijk}$ (which does not contain 
$k^{\mu}$),
the part (III): those including $B_{ij}$ (which does not contain 
$k^{\mu}$), 
and the part (IV): the rest part.
We will compare each part of $S^{''WZ}$ separately
with that of $S^{WZ}_{KK8A}$,
which in our notation takes the 
form\cite{exotic}:\footnote{In ref.\cite{exotic},
the factor $1/2^{6}$ of the first term of the last line is omitted.
Note that the notation in ref.\cite{exotic} is changed as
$C^{(3)}\to -C^{(3)},\  B\to -B$ and $A_{i} \to -A_{i}$.} 
\beqa
S_{KK8A}^{WZ}&=&\int d^{8} \xi \ \epsilon^{(8)}\cdot
[-\frac{1}{8!}(i_{k}N^{(9)})
-\frac{1}{7!}(i_{k}N^{(8)})\partial \omega^{(0)}\nonumber\\
&+&
\frac{1}{2\cdot 6!}(i_{k}N^{(7)})\{{\cal K'}^{(2)}
-2(i_{k}B^{(2)})DX C^{(1)}\}
-\frac{1}{2\cdot 5!}(i_{k}B^{(6)}){\cal K}^{(1)}{\cal K}^{(2)}
\nonumber\\
&-&\frac{1}{2^{3}\cdot 4!}(i_{k}C^{(5)})
\{ {\cal K}^{(2)} +4{\cal K}^{(1)}(DX C^{(1)})\}{\cal K}^{(2)}
\nonumber\\
&-&\frac{1}{24\cdot 3!}\widetilde{C^{(3)}}
\{2((i_{k}C^{(3)})(i_{k}B)-3{\cal K}^{(1)}\partial \omega^{(1)})
{\cal K}^{(2)}-(i_{k}C^{(3)})^{2}\partial \omega^{(0)}\}\nonumber\\
& + &\frac{1}{12\cdot 3!}\{2C^{(3)}(i_{k}B)+3(i_{k}C^{(3)})B\}
(i_{k}C^{(3)})(DX C^{(1)})\partial \omega^{(0)}\nonumber\\
&-&\frac{1}{48}\widetilde{B}\{(i_{k}C^{(3)})^{2}{\cal K}^{(2)}
-4(\partial \omega^{(1)})^{3}\}
+\frac{1}{3!}A\partial \omega^{(0)}(\partial \omega^{(1)})^{3}
\nonumber\\
&+&\frac{1}{2^{6}\cdot 3!}\frac{e^{2\phi}|k|^{-2}(i_{k}C^{(1)})}
{1+e^{2\phi}|k|^{-2}(i_{k}C^{(1)})^{2}}({\cal K}^{(2)})^{4}
+\frac{1}{7!}\partial \omega^{(7)}],\label{kk8a}
\eeqa
where $\epsilon\cdot T\equiv \epsilon^{i_{1}\cdots i_{8}}
T_{i_{1}\cdots i_{8}}$ for (products of) fields $T_{i_{1}\cdots
i_{8}}$ and 
${\cal K}^{(2)}={\cal K'}^{(2)}-2{\cal K}^{(1)}(DX C^{(1)})$.
$N^{(9)}$ is a 9-form gauge field introduced in ref.\cite{exotic},
which minimally couples to the KK-8A brane. 
We note that 
it holds $D_{i}X^{\mu}C^{(1)}_{\mu}
=\partial_{i} X^{\mu}
[C^{(1)}_{\mu}+|k|^{-2}k_{\mu}(i_{k}C^{(1)})]$.

First, we consider the part (I). 
That of ${\cal L''}^{WZ}$ comes from
the first three terms of 
the M-9-brane WZ action (\ref{m9action2}),
which give the following contribution:
\beqa
& &
\hat{\epsilon} 
\cdot [\frac{1}{9!}(i_{\hat{k}}\hat{C}^{(10)}) 
+\frac{1}{2\cdot 7!}
\{(i_{\hat{k}}\hat{N}^{(8)}) \hat{{\cal K'}}^{(2)}\}
+\frac{1}{2^{3}\cdot 5!}
(i_{\hat{k}}\hat{C}^{(6)})(\hat{{\cal K'}}^{(2)})^{2}]\ |
_{{\footnotesize Dimensional\  Reduction}}
\nonumber\\
&=&\epsilon^{(8)}\cdot [
-\frac{1}{8!}(i_{k}C^{'(9)}) 
-\frac{1}{7!} (i_{k}N^{(8)}){\cal K}^{(1)}
+\frac{1}{2\cdot 6!} (i_{k}N^{(7)}){\cal K'}^{(2)}
-\frac{1}{2\cdot 5!}(i_{k}B^{(6)}){\cal K'}^{(2)}{\cal K}^{(1)}
\nonumber\\
& &-\frac{1}{2^{3}\cdot 4!}(i_{k}C^{(5)})({\cal K'}^{(2)})^{2}
+\frac{1}{2^{3}\cdot 4!}\{3(i_{k}C^{(3)})B+2C^{(3)}(i_{k}B)\}
({\cal K'}^{(2)})^{2}] 
\label{drofc6}
\eeqa
where $\hat{\epsilon}\cdot \hat{T}\equiv 
\hat{\epsilon}^{\hat{i_{1}}\cdots\hat{i_{9}}}
\hat{T}_{\hat{i_{1}}\cdots\hat{i_{9}}}$ 
and we identify $\epsilon^{(8) i_{1}\cdots i_{8}}\equiv 
\hat{\epsilon}^{i_{1}\cdots i_{8}8}$.
So, the first five terms of the right hand side of (\ref{drofc6})
is the part (I), which we denote as ${\cal L''}^{WZ}|_{(I)}$.
On the other hand, 
the part (I) of ${\cal L}^{WZ}_{KK8A}$ 
corresponds to
the upper three lines of (\ref{kk8a}).
Subtracting ${\cal L''}^{WZ}|_{(I)}$
from the upper three lines of (\ref{kk8a}), 
we have the difference of the two\footnote{In deriving 
(\ref{difference1}),
the following identities are useful: 
${\cal K}^{(1)}({\cal K}^{(2)}-{\cal K'}^{(2)})\propto 
{\cal K}^{(1)}\wedge
{\cal K}^{(1)}=0$ and $\{{\cal K}^{(2)}+4{\cal K}^{(1)}DXC^{(1)}\}
{\cal K}^{(2)}=({\cal K'}^{(2)})^{2}$.}
\beqa
\{{\cal L}_{KK8A}^{WZ}-{\cal L''}^{WZ}\}|_{(I)}= 
\frac{1}{8!}\epsilon^{(8)} \cdot [(i_{k}C^{'(9)})-(i_{k}N^{(9)})+
8(i_{k}N^{(8)})(i_{k}B)\nonumber\\
-56(i_{k}N^{(7)})(i_{k}B)  
\{C^{(1)}+|k|^{-2}k_{\mu}\partial X^{\mu}(i_{k}C^{(1)})
].\label{difference1}
\eeqa
The right hand side of (\ref{difference1}) consists only of some
products of target-space fields, except for the factor $\partial X$.
Thus, the difference can be absorbed in the field 
redefinition relation of $C^{'(9)}$ and $N^{(9)}$.

Let us next consider the part (II) of ${\cal L''}^{WZ}$, 
which comes from the forth term of (\ref{m9action2}) and the last 
term of (\ref{drofc6}), as 
\beqa
{\cal L''}^{WZ}|_{(II)}=\frac{1}{2^{3}\cdot 3^{2}}\epsilon^{(8)}
\cdot [C^{(3)} 
\{6(\partial \omega^{(1)})^{2}\partial \omega^{(0)}
+6(i_{k}B)(\partial \omega^{(1)})^{2}
-3(i_{k}C^{(3)})\partial \omega^{(1)}\partial \omega^{(0)}
\nonumber\\
-5(i_{k}B) (i_{k}C^{(3)})\partial \omega^{(1)}
+\frac{1}{2}(i_{k}C^{(3)})^{2}\partial \omega^{(0)}
+\frac{3}{2}(i_{k}B)(i_{k}C^{(3)})^{2}\label{c3term}
\}].\ \ \ 
\eeqa
The part (II) of ${\cal L}^{WZ}_{KK8A}$ comes from the forth
line and the first term of the fifth line
in (\ref{kk8a}). 
The difference of the two is
\beqa
\{{\cal L}_{KK8A}^{WZ}-{\cal L''}^{WZ}\}|_{(II)}\sim
-\frac{1}{2^{4}\cdot 3^{2}}\epsilon^{(8)}\cdot 
[C^{(3)}(i_{k}C^{(3)})^{2}
(i_{k}B)],
\eeqa
which also can be absorbed in the redefinition of the 9-forms.
The part (III) is also discussed in the same way.
The part (III) of ${\cal L''}^{WZ}$ comes from the
forth term of (\ref{m9action2}) and the sixth 
term of (\ref{drofc6}): 
\beqa
{\cal L''}^{WZ}|_{(III)}
=\frac{1}{2^{4}\cdot 3!}\epsilon^{(8)}\cdot [B\{
8(\partial \omega^{(1)})^{3}
-4(i_{k}C^{(3)})^{2}\partial \omega^{(1)}
+(i_{k}C^{(3)})^{3}
\}].
\eeqa
The part (III) of $S^{WZ}_{KK8A}$ comes from
the second term of the fifth line and
the first term of the sixth line in (\ref{kk8a}).
The difference of this part is 
\beqa
\{{\cal L}_{KK8A}^{WZ}-{\cal L''}^{WZ}\}|_{(III)}&=&
\epsilon^{(8)}\cdot
\frac{1}{2^{4}\cdot 3!}B [ (i_{k}C^{(3)})^{3}\nonumber\\
&+&4(i_{k}C^{(3)})^{2}(i_{k}B)\{C^{(1)}
+|k|^{-2}k_{\mu}\partial X^{\mu}(i_{k}C^{(1)})\}]
\eeqa
which can also be absorbed in the same way. 

Finally, we consider
the part (IV), the rest part of the actions.
We can easily see that the last term of (\ref{m9action2}) gives the
last term of (\ref{kk8a}).
The sixth term $\frac{m}{5!}\hat{b}_{i_{1}}(\partial \hat{b})^{4}$
in (\ref{m9action2}) 
is the ``massive part'', i.e. the part which
arises when the background is a massive one. 
In fact, the massive part of the KK-8A brane action 
has not been discussed in ref.\cite{exotic}.
This means that (\ref{kk8a}) is not the action describing the KK-8A
brane in the most generic background.
The purpose of this subsection is 
to examine the correspondence of the {\it same part} of the two
actions.
Thus, in this subsection we set the mass parameter to be zero
($m=0$) and concentrate our discussion on
the massless part of the two actions.
We will argue the validity of this discussion 
under the setting $m=0$ and
discuss the massive part of the action in the final section.

The rest of the part (IV) of ${\cal L''}^{WZ}$ comes from
the fifth terms of (\ref{m9action2}):
\beqa
{\cal L''}^{WZ}|_{(IV)}&= & \frac{1}{2\cdot 4!}
\epsilon^{(8)}\cdot A [
{\cal K}^{(1)} ({\cal K}^{'(2)})^{3}]\nonumber\\
&+&\frac{1}{2^{4}\cdot 4!}\frac{e^{2\phi}|k|^{-2}(i_{k}C^{(1)})}
{1+e^{2\phi}|k|^{-2}(i_{k}C^{(1)})^{2}}
\epsilon^{(8)}\cdot [({\cal K}^{'(2)})^{4}
+8(DX C^{(1)}){\cal K}^{(1)}({\cal K}^{'(2)})^{3}\label{redofa}
].\ \ \ \ \ 
\eeqa
To derive this, we use the second expression of (\ref{ai2}).
The second term of (\ref{redofa}) exactly reproduces the last line
of (\ref{kk8a}).
The rest of the contribution of ${\cal L}_{KK8A}^{WZ}$ comes from 
the forth and the sixth lines of (\ref{kk8a})
since $\widetilde{T^{(r)}} 
=T^{(r)} 
-r\cdot A 
(i_{k}T^{(r)})$.
After a little calculation,
we obtain the difference of this part as
\beqa
\{{\cal L}_{KK8A}^{WZ}-{\cal L''}^{WZ}\}|_{(IV)}
=\frac{1}{48}\epsilon^{(8)}\cdot [|k|^{-2}k_{\mu}\partial X^{\mu}
(i_{k}B)(i_{k}C^{(3)})^{3}].
\eeqa
which can also be absorbed in the same way as the above.
(We note that $A_{i}=-|k|^{-2}k_{\mu}\partial_{i} X^{\mu}$.)

Therefore, putting together all the differences 
presented above, 
we can see that the two actions are equivalent 
if the following field redefinition relation of the 9-forms
holds:
\beqa
(i_{k}C^{'(9)})&=&(i_{k}N^{(9)})-8(i_{k}N^{(8)})(i_{k}B)
+56(i_{k}N^{(7)})(i_{k}B)\{C^{(1)}+|k|^{-2}k_{(\mu)}(i_{k}C^{(3)})\}
\nonumber\\
&+ &280 C^{(3)}(i_{k}C^{(3)})^{2}(i_{k}B)\nonumber\\
&-&420B [(i_{k}C^{(3)})^{3} 
+4(i_{k}C^{(3)})^{2}(i_{k}B)\{C^{(1)}
+|k|^{-2}k_{(\mu)}(i_{k}C^{(1)})\}]\nonumber\\
&-&840 |k|^{-2}k_{(\mu)}(i_{k}C^{(3)})^{3}(i_{k}B).
\label{redef9fm}
\eeqa
Thus, all we have to do now is only 
to show that (\ref{redef9fm}) really holds.
Actually,
correspondence of gauge transformations of the fields
is the only criterion in this theory
to discuss the consistency the field redefinition relation;
no other ingredients to judge the consistency,
such as supersymmetry, 
have not been discussed in this theory.\footnote{In addition,
although the 9-form $N^{(9)}$ is introduced in ref.\cite{exotic}
as a field to couple to the KK-8A brane,
the field strength of $N^{(9)}$ 
or the duality relation which it should satisfy 
have not been given. Thus, there is no other way to discuss 
the consistency of (\ref{redef9fm}). Conversely, 
we are able to determine
the field strength of $N^{(9)}$ through (\ref{redef9fm}),
by deriving the field strength of $C^{'(9)}$ 
from the field strength of $\hat{C}^{(10)}$
via dimensional reduction.}
To be concrete, we prove 
that the gauge transformations of
both sides of (\ref{redef9fm}) agree with each other.

First, we discuss the gauge transformation of
the right hand side (r.h.s) of (\ref{redef9fm}).
In the massless case $m=0$, the gauge transformations of fields are
defined as\cite{berg4}\cite{lozkkm}\cite{exotic} 
\beqa
\delta (i_{k}N^{(9)})&=&8\partial(i_{k}\Omega^{(8)})
-168\partial(i_{k}\Omega^{(6)})
[i_{k}C^{(3)}+2(i_{k}B)\{C^{(1)}+|k|^{-2}k_{(\mu)}(i_{k}C^{(1)})\}]
\nonumber\\
&+&840\partial(i_{k}\Lambda^{(4)})(i_{k}C^{(3)})
[i_{k}C^{(3)}+4(i_{k}B)\{C^{(1)}+|k|^{-2}k_{(\mu)}(i_{k}C^{(1)})\}]
\nonumber\\
&+&2520\partial(i_{k}\Lambda^{(2)})(i_{k}C^{(3)})
[(i_{k}C^{(3)})\{B-2|k|^{-2}k_{(\mu)}(i_{k}B)\}\nonumber\\
& &\ \ \ \ \ \ \ \ \ \ \ \ \ \ \ \ \ \ \ \ \ \ \ \ \ \ 
+4B(i_{k}B)\{C^{(1)}+|k|^{-2}k_{(\mu)}(i_{k}C^{(1)})\}]
\nonumber\\
&-&8\partial(i_{k}\Lambda^{(1)})[i_{k}N^{(8)}
+7(i_{k}N^{(7)})\{C^{(1)}+|k|^{-2}k_{(\mu)}(i_{k}C^{(1)})\}
\nonumber\\
& &\ \ \ \ \ \ \ \ \ \ \ \ \ \ \ 
-35(i_{k}C^{(3)})^{2}\{C^{(3)}+3|k|^{-2}k_{(\mu)}(i_{k}C^{(3)})\}
\nonumber\\
&-&70(i_{k}C^{(3)})\{2C^{(3)}(i_{k}B)+3(i_{k}C^{(3)})B \}
\{C^{(1)}+|k|^{-2}k_{(\mu)}(i_{k}C^{(1)})\}]\\
\delta (i_{k}N^{(8)})&=&7\partial(i_{k}\Omega^{(7)})
+105\partial(i_{k}\Lambda^{(5)})(i_{k}C^{(3)})-210\partial\Lambda^{(2)}
(i_{k}C^{(3)})^{2}\nonumber\\
&+&140\partial(i_{k}\Lambda^{(2)})C^{(3)}(i_{k}C^{(3)})
-7\partial\Lambda^{(0)}(i_{k}N^{(7)})\\
\delta N^{(7)}&=&6\partial(i_{k}\Omega^{(6)})
+30\partial(i_{k}\Lambda^{(5)})(i_{k}B)
-60\partial(i_{k}\Lambda^{(4)})(i_{k}C^{(3)})\nonumber\\
&-&120\partial\Lambda^{(2)}(i_{k}C^{(3)})(i_{k}B)
+20\partial(i_{k}\Lambda^{(2)})\{2C^{(3)}(i_{k}B)-3(i_{k}C^{(3)})B\}
\nonumber\\
& &-20\partial(i_{k}\Lambda^{(1)})C^{(3)}(i_{k}C^{(3)})
\\
\delta \tilde{B}^{(6)}&=&6\partial\Lambda^{(5)}
-30\partial\Lambda^{(2)}C^{(3)}+6\partial\Lambda^{(0)}
\{C^{(5)}-5C^{(3)}B\}
\\
\delta C^{(5)}&=&5\partial\Lambda^{(4)}+30\partial\Lambda^{(2)}B
+15\partial\Lambda^{(0)}BB
\\
\delta C^{(3)}&=&3\partial\Lambda^{(2)}+3\partial\Lambda^{(0)}B
\\
\delta B&=&2\partial\Lambda^{(1)}
\\
\delta C^{(1)}&=&\partial\Lambda^{(0)}
\eeqa
where $\Omega^{(r)}$ and $\Lambda^{(r)}$
are the r-form gauge parameters associated with the (r+1)-form gauge
potentials, respectively.\footnote{Note that the notation in 
refs.\cite{lozkkm}\cite{exotic}
is changed as
$\Lambda^{(2)}\to -\Lambda^{(2)}, \ 
\Lambda \to -\Lambda^{(1)}, \ \Sigma^{(6)} \to \Omega^{(6)}$ and
$\tilde{\Lambda} \to \Lambda^{(5)}$
(in addition to $C^{(3)}\to -C^{(3)},\  B\to -B$).} 
We note that
these parameters, except for the $\Omega^{(8)}$ which is associated
with $N^{(9)}$,
are related to those in 11D as\footnote{
$\delta C^{(1)}$ corresponds to the coordinate transformation
in the direction parametrized by $\hat{x}^{8}$. We do not discuss
the transformation with respect to $\Lambda^{(0)}$ here.}

\beqa
\left. 
\begin{array}{cc}
\left\{
\begin{array}{cc}
(i_{\hat{k}}\hat{\Omega}^{(7)})_{\mu_{1}\cdots\mu_{6}}
= &-(i_{k}\Omega^{(7)})_{\mu_{1}\cdots\mu_{6}}\\
(i_{\hat{k}}\hat{\Omega}^{(7)})_{\mu_{1}\cdots \mu_{5}8}
= &(i_{k}\Omega^{(6)})_{\mu_{1}\cdots \mu_{5}}
\end{array}
\right.
&
\left\{
\begin{array}{cc}
(i_{\hat{k}}\hat{\chi}^{(5)})_{\mu_{1}\cdots\mu_{4}}
= &-(i_{k}\Lambda^{(5)})_{\mu_{1}\cdots\mu_{4}}\\
(i_{\hat{k}}\hat{\chi}^{(5)})_{\mu_{1}\cdots \mu_{3}8}
= &-(i_{k}\Lambda^{(4)})_{\mu_{1}\cdots \mu_{3}}\\
\end{array}
\right.
\end{array}
\right. 
\nonumber\\
\left\{
\begin{array}{cc}
\hat{\chi}_{\mu\nu}= &\Lambda^{(2)}_{\mu\nu}\\
\hat{\chi}_{\mu 8}= &\Lambda^{(1)}_{\mu}.
\end{array}
\right.\label{gaugepararelof11and10}
\eeqa
After a bit lengthy calculation,
the {\it massless part} of the transformation of the r.h.s of
(\ref{redef9fm}) is written as
\beqa
\delta ({\rm r.h.s. \ of \ 
(\ref{redef9fm})})&=&8\partial (i_{k}\Omega^{(8)})
-56(i_{k}\Omega^{(7)})(i_{k}B)-168(i_{k}\Omega^{(6)})(i_{k}C^{(3)})
\nonumber\\
&-&
840\partial(i_{k}\Lambda^{(5)})(i_{k}C^{(3)})(i_{k}B)
+840 \partial(i_{k}\Lambda^{(4)})(i_{k}C^{(3)})^{2}\nonumber\\
&+&2520\partial \Lambda^{(2)}(i_{k}C^{(3)})^{2}(i_{k}B)
+840\partial \Lambda^{(1)}(i_{k}C^{(3)})^{3}.\label{gaugetrofc'}
\eeqa
On the other hand, the massless part of the transformation of
$C^{'(9)}$ is obtained from (\ref{c10gaugetr}) via the 
dimensional reduction.
Using (\ref{gaugepararelof11and10}), and
if we identify $\hat{\Omega}^{(9)}$ with $\Omega^{(8)}$ as
\beqa
(i_{k}\Omega^{(8)})=-(i_{\hat{k}}\hat{\Omega}^{(9)})
_{\mu_{1}\cdots \mu_{7}8},
\eeqa
the gauge transformation of $C^{'(9)}$ completely agrees 
with the r.h.s of (\ref{gaugetrofc'}) !
Thus, we conclude that the WZ action of the KK-8A brane (which
we regard as a kind of D-8-brane)
is certainly obtained from that of the M-9-brane
on this dimensional reduction.
Therefore, on the double dimensional reduction,
the relations of branes in Figure 1 is consistent
from WVEA point of view.



\subsection{The dimensional reduction along the transverse
direction}

Finally, we discuss the dimensional reduction of $S_{M9}^{WZ}$
along the single direction transverse to the M-9-brane, 
and show that the NS-9A brane WZ action $S_{NS9A}^{WZ}$
is derived on this reduction.

In this case, we split 
$\hat{x}^{\hat{\mu}}$ 
into $(x^{\mu},y)$ ($\mu=0,1,\cdots,8,z$), 
and dimensionally reduce the 11D fields along $y$.
Naively, the SUGRA describing the bulk
might be expected to be the same as the KK-8A brane case,
since so are
the relations of the target-space fields in 11D 
with those in 10D except for the replacement $z$ or $8\to y$.
However, a certain truncation procedure is needed in this case
because
the obtained brane
is a 9-brane in 10 dimensions, i.e. a spacetime-filling brane.
The famous example is the case of D-9-branes 
in IIB theory\cite{pol2};
the full field content of IIB theory is
$\{g_{\mu\nu},\phi,B,B^{(10)},C^{(0)},C^{(2)},C^{(4)},C^{(10)}\}$,
but due to the conservation law of the charge associated with 
$C^{(10)}$, the number of D-9-branes is constrained to be 32, and 
that orientifolding by the worldsheet parity is needed.
In this orientifold construction, the IIB multiplet is 
truncated to that of the N=1 SUGRA with the condition $\{
B=C^{(0)}=C^{(4)}=B^{(10)}=0 \}$, and Type I string theory arises.
Then, starting from this case and using S- and T-duality, 
the authors of ref.\cite{sptfilling}
have argued that if the NS-9A branes exist,
their number is also constrained to be 32, and that
the following truncation condition is imposed on 
the RR fields of the IIA SUGRA: 
\beqa
C^{(1)}=C^{(3)}=0\label{trunc1}\\
C^{(9)}=0.\label{trunc2}
\eeqa
In particular, the last condition is derived via T-duality
from the condition that the IIB 10-form $C^{(10)}$ should vanish.
However, this 10D theory still has a Killing isometry with the vector
$k^{\mu}\equiv \hat{k}^{\mu}$. (Note that $\hat{k}^{y}=0$.)
Thus, precisely speaking, 
the 9-form potential which should arise as
a T-dual of the IIB 10-form $C^{(10)}$ in this case
is 
not the usual 9-form $C^{(9)}$
but the 9-form $C^{'(9)}$.
Therefore, instead of (\ref{trunc2}),
we set the truncation condition: 
\beqa
C^{'(9)}=0.\label{trunc3}
\eeqa
Moreover, this condition, together with
the duality relation (\ref{dualityofc9'})
implies a further condition $m=0$.

On the other hand,
the truncation conditions on 
$i_{k}N^{(8)}$ and $i_{k}N^{(7)}$ have not been
discussed before, but in this case 
we can infer it from the latter of the two duality relations 
(\ref{duality1}).
Dimensionally reducing the relation along $y$
and substituting the truncation conditions (\ref{trunc1}) and
(\ref{trunc2}) for it,
we have the condition $\partial (i_{k}N^{(8)})=0$, 
which essentially means $i_{k}N^{(8)}=0$. 
We note that no truncation condition is imposed on
$i_{k}N^{(7)}$.

As for the worldvolume fields,
a vector field and an 8-form $\omega^{(8)}_{i_{1}\cdots i_{8}}$
appears,  coming from $\hat{b}_{i}$ and the 8-form 
$\hat{\omega}^{(8)}_{i_{1}\cdots i_{8}}$, respectively.
In addition, a scalar field $Y(\xi)$ arises
from the embedding of $\xi^{i}$ in the transverse coordinate 
$\hat{x}^{y}=y$.
We identify 
$\hat{b}_{i} =d^{(1)}_{i}$ for $i=0,\cdots,8$
and $Y(\xi)=c^{(0)}$ where
$d^{(1)}_{i}$ is
a vector field and $c^{(0)}$ is a scalar field, both defined in 
ref.\cite{sptfilling}.
We note that after the truncation,
the dimensional reduction of $\hat{A}_{i}$ is deduced from 
(\ref{ai2}) as 
$\hat{A}_{i}=A_{i}\equiv -|k|^{-2}\partial_{i}X^{\mu}k_{\mu}$.

Then, the WZ action of the 9-brane $S^{'''WZ}$ 
obtained from that of the M-9-brane 
(\ref{m9action2}) is written as
\beqa
S^{'''WZ}&=&\int d^{9}\xi \epsilon^{i_{1}\cdots i_{9}} [ 
-\frac{1}{9!}(i_{k}B^{(10)})
+\frac{1}{6!}(i_{k}N^{(7)})\partial c^{(0)} \partial
d^{(1)}\nonumber\\
& &-\frac{1}{2\cdot 5!}(i_{k}B^{(6)})
\{(\partial d^{(1)})^{2}+\partial d^{(1)}(i_{k}B)\partial c^{(0)}\}
+\frac{1}{2\cdot 3!} B(\partial d^{(1)})^{3}\partial
c^{(0)}\nonumber\\
& &+\frac{1}{3!}A(i_{k}B)(\partial d^{(1)})^{3}\partial
c^{(0)}+\frac{1}{4!}A_{i}(\partial d^{(1)})^{4}
+\frac{1}{8!}\partial \omega^{(8)}
]_{i_{1}\cdots i_{9}}.\label{NS9Aaction}
\eeqa
This is exactly the same as the WZ action of the NS-9A brane given in 
ref.\cite{sptfilling} !\footnote{To make the 
complete correspondence
with the action of ref.\cite{sptfilling},
we have only change our notation as
$C^{(3)}\to -C^{(3)},\  B\to -B$ and $A_{i} \to -A_{i}$}
Thus, we conclude that on the direct dimensional reduction of the
M-9-brane, 
the relations in Fig.1 is also consistent from WVEA point of view.

The explicit correspondence of the terms in the actions is as follows:
Each of the first three terms of the M-9-brane WZ action 
(\ref{m9action2}) only gives each of 
the first three terms of the NS-9A brane
WZ action (\ref{NS9Aaction}), respectively.
The fourth term (including $\hat{C}$) in (\ref{m9action2})
gives the fourth term (including $B$) in (\ref{NS9Aaction}), and
the fifth term (including $\hat{A}_{i}$) in (\ref{m9action2})
gives the fifth and the sixth terms (including $A_{i}$) 
in (\ref{NS9Aaction}).
The last term in (\ref{m9action2}) 
gives the last term 
in (\ref{NS9Aaction}), of course.
All the other terms vanish, mainly due to the truncation conditions.
We note that it holds
$(i_{k}B)^{2}=(\partial c^{(0)})^{2}=0$ in the action.


\section{ 
Summary and discussion}

In this paper we have 
shown 
that on dimensional reductions along three
different directions, the Wess-Zumino action of the M-9-brane
respectively gives those of the D-8-brane, 
the KK-8A brane (which we regard as a D-8-brane
in some special background)
and the NS-9A 9-brane,  the last two of which were obtained
via dualities. 
Therefore, we conclude 
that the relation of p-branes for $p\ge 8$, or
of the M-9-brane with the other branes, proposed
previously\cite{hullalg}\cite{meessen1}\cite{sptfilling}\cite{eyras1}
(and represented in Fig.1) is consistent
from the viewpoint of worldvolme actions.


Now, we discuss the massive part of the KK-8A brane WZ action.
In section 3.2  we have set $m=0$ and 
discussed the massless part of the actions, 
since only massless part of $S^{WZ}_{KK8A}$ 
has been discussed in ref.\cite{exotic}.
However, 
the mass parameter $m$ corresponds nearly to 
the field strength dual of the 9-form
$C^{'(9)}$ (and hence $N^{(9)}$) (see (\ref{dualityofc9})), so
setting the mass parameter to be zero 
essentially implies some trivial configuration of the background
9-form potential $C^{'(9)}$ (or $N^{(9)}$).
Thus, the massive part should also be taken into account
if one want the action of the KK-8A brane in a more general 
background.
So, as another result of this paper, 
we propose the massive part of $S_{KK8A}^{WZ}$ which was not
obtained previously;
It is derived from the dimensional reduction of
$\frac{m}{5!}\hat{b}_{i_{1}}
(\partial \hat{b})^{4}$, the massive part of $S_{M9}^{WZ}$, as
\beqa
S_{KK8A}^{WZ}|_{{\rm massive}}=\int d^{9}\xi\epsilon^{(8)i_{1}\cdots
i_{8}}
\frac{m}{5!}\{
\omega^{(0)}(\partial \omega^{(1)})^{4}-4(\partial \omega^{(0)})
\omega^{(1)}(\partial \omega^{(1)})^{3}\}_{i_{1}\cdots
i_{8}}.\label{kk8amassive}
\eeqa
On the other hand, 
the action of the KK-8A brane (\ref{kk8a})
is derived from that of 
the D-7-brane via S- and T-dualities defined in ref.\cite{exotic},
but the duality relations of target-space fields discussed in 
ref.\cite{exotic} are only the massless parts of them.
They should be generalized to those in some massive
background (like the usual case done 
in ref.\cite{berg3}\cite{massDbr1}),
so that the massive part of the KK8A brane WZ action
(\ref{kk8amassive}) is reproduced,
but we do not discuss them further here.

Here we would like to note the following two things:
One is that trivial configuration of $C^{'(9)}$
does not mean the inconsistency of the action $S^{WZ}_{KK8A}$
but merely implies some loss of generality of the background.
(A similar case happens if one consider the worldvolume action
of a D-8-brane in a {\it massless} background.)
Another is that
the consistency check of the
field redefinition relation (\ref{redef9fm}) we have made in 
section 3.2 is not a trivial one.
The reason is that
even if the configuration of $C^{'(9)}$ is trivial,
the gauge transformations of $C^{'(9)}$
and the other fields 
do not become trivial; They are not affected
except for the setting $m=0$.



Finally, we would like to give a comment on 
the relation of the existence of
the KK-8A brane with spacetime superalgebras. 
In ref.\cite{eyras1}\cite{exotic}, it is argued that
the KK-8A brane is a brane not predicted by the IIA spacetime superalgebra.
However, we do not agree with this argument.
Our objection is based on the fact that the existence of 
the M-9-brane, the 11D origin of the
KK-8A brane, is suggested by the 11D superalgebra:
Suppose we denote p-form central charges of superalgebras
as $Z^{(p)}_{\mu_{1}\cdots\mu_{p}}$.
Then, in the massive 11D theory where an isometry direction 
parametrized by $z$ is assumed, 
the existence of an M-9-brane extending to 
the directions of $\hat{x}^{1},\cdots \hat{x}^{8}$ and $z$
corresponds to a non-vanishing central charge $\hat{Z}^{(2)}_{09}
=\hat{\widetilde{Z}}^{(9)}_{01\cdots 8z}$ 
of the 11D superalgebra\cite{hullalg},
where $\widetilde{Z}^{(D-p)}_{\mu_{1}\cdots\mu_{D-p}}$ 
indicates the dual central charge of $Z^{(p)}$ in a D-dimensional
superalgebra.
If the M-9-brane 
is dimensionally reduced along $z$, it gives a D-8-brane 
and the D-8-brane has a corresponding charge
$\widetilde{Z}_{01\cdots 8}\equiv
\hat{\widetilde{Z}}^{(9)}_{01\cdots 8z}$ of the D=10 superalgebra
of the usual massive IIA theory.
In the same way, if the M-9-brane
is dimensionally reduced along $x^{8}$,
it gives a KK-8A brane (as a kind of D-8-brane),
and the KK-8A brane should have
a corresponding charge $\widetilde{Z}_{01\cdots 7z}'\equiv
-\hat{\widetilde{Z}}^{(9)}_{01\cdots 8z}$
of the D=10 superalgebra of 
{\it another} massive IIA theory with an isometry direction.
(The prime ' of $\widetilde{Z}$ implies ``another''.)
Therefore, we argue that, at least only in the case of KK-8A brane 
the information of
the existence of the brane is 
included in the D=10 spacetime superalgebra.

\parbigskipn

\noindent
{\large \bf Acknowledgment}

\parbigskipn
We would like to thank Taro Tani and Tunehide Kuroki for fruitful
discussions and encouragement. 
We would also like to thank Shinya Tamura for some comments.
We are grateful to Prof. Eric Bergshoeff for useful comments
and Prof. Yolanda Lozano for useful comments via email.

\parbigskipn



\end {document}